\documentstyle[preprint,aps,prc]{revtex}
\newcounter{fig}
\newcounter{ref1}
\newcounter{ref2}
\begin{document}

\draft

\title{Microscopic analysis of quadrupole collective motion
in Cr-Fe nuclei\\
\Roman{ref2}. Doorway nature of mixed-symmetry states}

\author{Hitoshi Nakada}
\address{Department of Physics, Faculty of Science, Chiba University\\
Yayoi-cho 1-33, Inage-ku, Chiba 263, Japan}
\author{Takaharu Otsuka}
\address{Department of Physics, Faculty of Science, 
University of Tokyo\\
Hongo 7-3-1, Bunkyo-ku, Tokyo 113, Japan}

\date{\today}

\maketitle

\begin{abstract}
The mixed-symmetry collective modes are investigated in Cr-Fe nuclei,
by analyzing the realistic shell-model wavefunctions
via the $H^n$-cooling method.
It is clarified that the relatively low-lying mixed-symmetry states
behave like doorway states.
For these nearly spherical nuclei,
the lowest mixed-symmetry state is shown to have $J^P=2^+$.
An indication of the mixed-symmetry $3^+$ state is obtained.
The sequence of the mixed-symmetry $2^+$, $1^+$ and $3^+$ levels
and its nucleus-dependence are discussed.
Calculated $M1$ and $M3$ transitions in the low-energy region
suggest that the mixed-symmetry $1^+$ and $3^+$ components
are detectable.
We investigate the $B(M1)$ distribution in a wider energy range,
without breaking the isospin quantum number.
It is confirmed that the mixed-symmetry $1^+$ component
is well separated from the peak of the spin excitation.
The isospin-raising component has a peak,
separated well from the isospin-conserving one.
The orbital angular-momentum contributes destructively
to the spin excitations.
\end{abstract}



\section{Introduction}
\label{sec:intro}

While the mixed-symmetry (MS) states
of the proton-neutron interacting boson model (IBM-2)
have been studied for more than a decade,
there still remain various aspects to be explored
on this type of collective mode.
The $1^+$ states discovered around $E_x\simeq 3$~MeV
in rotational nuclei\cite{ref:scissors2}
correspond to a MS state (scissors mode),
which has been predicted by the IBM-2\cite{ref:scissors}
as well as by other models\cite{ref:scissors1}.
Significant fragmentation of the $B(M1)$ strength,
however, has been observed.
The MS $2^+$ state has remained much less investigated so far,
though this $2^+$ state may be lower
than any other MS states in spherical nuclei\cite{ref:predict}.
In terms of the geometrical picture,
the MS $2^+$ state in a spherical nucleus is interpreted
as the quadrupole oscillation out of phase
between protons and neutrons\cite{ref:iv-qosc}.
Experimental studies in the Cr-Fe region
have suggested the presence and rather weak fragmentation
of the MS $2^+$ state\cite{ref:mixsym2}.
We have reported a realistic shell-model result
on this state in $^{56}$Fe\cite{ref:NOS91}.

We shall study, in this paper,
the MS states of Cr-Fe nuclei in more detail.
These nuclei provide us with a precious clue
to understanding some basic features
of the nuclear quadrupole collective motion
from microscopic standpoints.
On one side, salient quadrupole collectivity
can be formed in these nuclei.
On the other side, these nuclei are accessible
by a realistic shell-model calculation\cite{ref:NSO94},
which becomes prohibitively difficult in heavier nuclei.
In order to discuss quadrupole collective modes
based on the shell model,
the $H^n$-cooling method (H$^n$CM),
which has been proposed in the previous paper\cite{ref:no.1},
will be used.
The wavefunctions of the MS states
have been generated explicitly with the renormalization
based on the H$^n$CM.
It is straightforward to compare these wavefunctions
with those of the realistic shell model.
We thus investigate some properties
of the relatively low-lying MS states
in connection to the realistic shell-model calculation.

\section{Shell-model calculation and renormalization}
\label{sec:shell}

It has been shown that a large-scale shell-model calculation
with a realistic interaction derived 
from the $G$-matrix\cite{ref:KBpf} is successful 
in the middle {\em pf}-shell region\cite{ref:NOS91,ref:NSO94}.
The $k\leq 2$ model space has been assumed,
where $k$ denotes the number of nucleons
excited from $0f_{7/2}$ to the upper {\em pf}-shell orbits.
For even-even nuclei, the energy levels with $E_x<4$~MeV
are described to a good accuracy,
with typical deviation of 0.3~MeV\cite{ref:NSO94}.
The large configuration space is crucial to reproduce the levels
in the energy region of $3<E_x<4$~MeV,
where the MS states likely appear.
The energy levels of $^{58}$Fe and $^{56}$Cr are also shown
in Ref.\cite{ref:no.1}.

We shall investigate the MS quadrupole collective modes,
based on this realistic shell-model result.
In the OAI mapping\cite{ref:OAI}, a correspondence is postulated
between the IBM-2 states and the fermion states comprised 
of the $SD$-pairs.
The realistic shell-model hamiltonian, however, 
contains various correlations,
which are not carried by the naive $S$- and $D$-pairs.
The total probability of the $SD$-components
is substantially smaller than unity
even in the shell-model $0^+_1$ and $2_1^+$ states.
In order to incorporate effects of the non-$SD$ components,
we renormalize the wavefunctions of the $SD$-states.
The leakage out of the $SD$ space occurs
because of the excitation from $0f_{7/2}$,
as well as of the admixture of like-nucleon pairs 
other than $S$ and $D$.
The excitation out of the $^{56}$Ni core
amounts to about $40\%$ in the ground states\cite{ref:NSO94},
whereas the influence of other nucleon pairs 
without the core excitation
is small in the $0_1^+$ and $2_1^+$ states.
This relatively large core-excitation probability requires a method
beyond the perturbation theory.
The H$^n$CM developed in the previous paper\cite{ref:no.1}
enables us to renormalize wavefunctions 
of the quadrupole collective states
and to discuss the MS states to the same extent
as the $0^+_1$ and $2_1^+$ states.

We shall briefly discuss here
the renormalization associated with the H$^n$CM briefly,
which has been introduced in Ref.\cite{ref:no.1}.
In the description assuming the $^{56}$Ni inert core,
the $^{56}$Fe nucleus has a pair of proton holes
and a pair of neutron valence particles.
The $S$- and $D$-pairs of protons are defined
as the $0^+$ and $2^+$ states of the $(0f_{7/2})^{-2}$ configuration,
while those of neutrons are collective $0^+$ and $2^+$ states
of the $(0f_{5/2}1p_{3/2}1p_{1/2})^2$ configuration.
The $SD$ space is then constructed
by these proton and neutron $S$- and $D$-pairs.
The proton-neutron symmetry is taken into consideration 
for the $SD$-states
through the $F$-spin values of their IBM-2 images\cite{ref:no.1}.
The states having $F=F_{\rm max}$
are called totally-symmetric states in the IBM-2,
while those with $F=F_{\rm max}-1$ are MS states.
Here $F_{\rm max}={1\over 2} N^{\rm B}=
{1\over 2} (N_\pi^{\rm B}+N_\nu^{\rm B})$,
with $N^{\rm B}$ denoting boson 
({\it i.e.}, like-nucleon pair) number.
The lowest-lying quadrupole collective states have been known
to be predominantly totally-symmetric states.
We will mainly deal with some lower-lying states
characterized by $F=F_{\rm max}-1$.
One might notice that, because $F_{\rm max}=1$,
the $F=F_{\rm max}-1$ states of $^{56}$Fe
turn out to be totally anti-symmetric.
However, they are classified as mixed-symmetry states in this article,
because they belong to the class of the $F=F_{\rm max}-1$ states.

In the H$^n$CM, the primary bases\cite{ref:no.1}
(denoted by $\Psi_\lambda^{(0)}$ with $\lambda=1,2,\cdots,l$)
are assumed at the beginning.
The bases consisting of the $SD$-pairs will be taken
as primary bases in the present case.
If $e^{-\beta H}$ acts on a primary basis,
each basis yields a superposition
of a number of exponentially-decaying components.
In the H$^n$CM with a given $n$,
we consider $n$ non-primary bases (denoted by $\phi_\lambda^{(1)}$,
$\phi_\lambda^{(2)}$, $\cdots$, $\phi_\lambda^{(n)}$)
for each $\Psi_\lambda^{(0)}$,
through the power-series expansion
of $e^{-\Delta\beta H}$ up to $O((\Delta\beta)^n)$.
The $\phi_\lambda^{(\nu)}$ basis is
generated from $H^\nu \left|\Psi_\lambda^{(0)}\right\rangle$.
In this procedure, an appropriate orthonormalization is applied.
We thus have the following primary and non-primary orthonormal bases,
\begin{equation} \begin{array}{cccc}
\Psi_1^{(0)},&\Psi_2^{(0)},&\cdots,&\Psi_l^{(0)},\\
\phi_1^{(1)},&\phi_2^{(1)},&\cdots,&\phi_l^{(1)},\\
\phi_1^{(2)},&\phi_2^{(2)},&\cdots,&\phi_l^{(2)},\\
\multicolumn{4}{c}{\cdots \cdots \cdots,}\\
\phi_1^{(n)},&\phi_2^{(n)},&\cdots,&\phi_l^{(n)}.
\end{array} \label{eq:ren-bases} \end{equation}
For each $\lambda$, we solve the eigenvalue problem within the space
$\Gamma_\lambda^{(n)}\equiv\left\{ \Psi_\lambda^{(0)},
\phi_\lambda^{(1)},\phi_\lambda^{(2)},\cdots,\phi_\lambda^{(n)} 
\right\}$,
having $(n+1)$ eigenvectors.
The lowest one is adopted as the renormalized basis 
$\Psi_\lambda^{(n)}$.
Once the renormalized bases are obtained,
the renormalized space is comprised of $\Psi_\lambda^{(n)}$'s.
This space closes up to $O((\Delta\beta)^n)$,
for the operation of $e^{-\Delta\beta H}$.

In the application of the H$^n$CM to the $SD$-space
of the Cr-Fe nuclei\cite{ref:no.1},
the angular-momentum conservation is treated exactly
in the whole procedure of the H$^n$CM,
by choosing the bases with good angular momentum 
for $\Psi_\lambda^{(0)}$'s.
Note that, because the proton and neutron pairs are formed
by different orbits,
all the $SD$-states have the same isospin for a given nucleus.
This situation is also maintained during the H$^n$CM.
Moreover, the proton-neutron symmetry ({\it i.e.}, $F$-spin),
which concerns our main interest of this paper,
is also taken into consideration in choosing the primary bases.

The effective hamiltonian within the renormalized $SD$ space
is given by the matrix elements 
$\left\langle\Psi_{\lambda'}^{(n)}\right|
H\left|\Psi_\lambda^{(n)}\right\rangle$.
The energy levels and corresponding wavefunctions
in the renormalized $SD$ space
are calculated by diagonalizing the matrix thus obtained.
Therefore each of these eigenfunctions is a linear combination
of $\Psi_\lambda^{(n)}$'s.
In the following sections, we shall inspect the eigenfunctions
within the renormalized $SD$ space,
focusing on the MS components.
The actual H$^n$CM procedure is carried out
for $n=2$ ({\it i.e.}, H$^2$CM) in this study.

As shown in Ref.\cite{ref:no.1},
the H$^2$CM provides us with remarkable improvement
on the $0^+_1$ and $2^+_1$ wavefunctions
from the unrenormalized ones,
in comparison with more complete shell-model results
(see Table~\Roman{ref1} of Ref.\cite{ref:no.1}).
The energy levels are also in good agreement,
between the H$^2$CM and the shell-model results
(see Figs.~1, 2, 5 and 6 of Ref.\cite{ref:no.1}).
A boson mapping, which is an extension of the OAI mapping,
has also been introduced in Ref.\cite{ref:no.1}.
The IBM-2 parameters have been evaluated via this mapping.
These parameters characterize some properties
of the quadrupole collective states.

\section{Energies of mixed-symmetry states with H$^2$CM}
\label{sec:energy}

We shall discuss, in this section, the structures of the eigenstates
obtained within the renormalized $SD$ space via the H$^2$CM,
for $^{56}$Fe, $^{54}$Cr, $^{58}$Fe and $^{56}$Cr.
The excitation energies and the $F$-spin contents 
of lower-lying states
are tabulated in Table~\ref{tab:F-prob}.
The $F$-spin values for the renormalized $SD$-states
are defined through the $F$-spin values 
of their IBM-2 images\cite{ref:no.1}.
It is seen that the lowest $0^+$ and $2^+$ states
are highly dominated by the symmetric components.
Though not all the states presented in Table~\ref{tab:F-prob}
have one-to-one correspondence to the shell-model eigenstates,
their identities seem to be maintained to a good extent,
as will be discussed in Sect.~\ref{sec:dist}.

\subsection{Mixed-symmetry $2^+$ state}
\label{subsec:2M}

In all the nuclei under investigation,
the lowest state dominated by the mixed-symmetry (MS) component 
has $J^P=2^+$;
it is the second $2^+$ state in the $SD$ space 
of $^{56}$Fe and $^{54}$Cr,
while in $^{58}$Fe and $^{56}$Cr it is the third $2^+$ state.
This is a clear indication,
derived from a microscopic and realistic standpoint,
that the MS $2^+$ state
is lower than any other MS states in these nuclei.
The excitation energy of these MS $2^+$ states are 3.1--3.7~MeV,
somewhat dependent on the nuclide.

Even though the $F$-spin is good for the $0^+_1$ and $2^+_1$ states,
it is not trivial how pure the $F$-spin is in the MS-dominant states,
because there may be some symmetric components
around the lowest-lying MS components.
As is shown in Table~\ref{tab:F-prob},
the $F$-spin purity of the second $2^+$ state of $^{56}$Fe 
is excellent.
This is consistent with the earlier analysis in Ref.\cite{ref:NOS91}.
Thus the $F$-spin plays a significant role
in classifying the collective states of $^{56}$Fe.

In contrast to $^{56}$Fe, there is a certain amount of mixing
between totally symmetric and MS components
in the other three nuclei.
Among them, $^{56}$Cr has a relatively good $F$-spin purity
for the MS-dominated $2^+$ state.
This can be discussed in association with the IBM-2 hamiltonian,
which has been derived via the extended OAI mapping 
in Ref.\cite{ref:no.1}.
The $F$-spin purity correlates mainly with the $\chi$ parameters
of the bosonic $\hat Q_\pi \cdot \hat Q_\nu$ interaction,
since $\epsilon_{d_\pi}$ and $\epsilon_{d_\nu}$
have rather close values to each other.
In $^{56}$Fe and $^{56}$Cr, the $F$-spin breaking 
due to the $\hat Q_\pi \cdot \hat Q_\nu$ interaction
is expected to be small, because $\chi_\pi \approx \chi_\nu$
(see Table~\Roman{ref2} of Ref.\cite{ref:no.1}).
It is noticed from Table~\ref{tab:F-prob}
that the influence of the $F=F_{\rm max}-2$ component,
which can be present in  $^{56}$Cr, is quite small.
Although the lack of the $|D_\pi^2;0^+\rangle$ component 
in the Cr nuclei
prohibits the $F$-spin from being pure,
this does not seriously influence the lower-lying states.

\subsection{Mixed-symmetry $1^+$ state}
\label{subsec:1M}

For the same reason as in the IBM-2,
we have no symmetric $1^+$ basis in the $SD$ space of any nucleus.
The lowest $1^+$ state in the $SD$ space
should therefore have $F=F_{\rm max}-1$
({\it i.e.}, MS state).
The H$^2$CM yields 3.7--4.8~MeV excitation energy
for the lowest collective $1^+$ state
of these Cr-Fe nuclei ($^{56}$Fe, $^{54}$Cr, $^{58}$Fe and $^{56}$Cr).

\subsection{Mixed-symmetry $3^+$ state}
\label{subsec:3M}

There is only a single $3^+$ basis in the $SD$ space 
of $^{56}$Fe, which is the MS one.
For $^{54}$Cr, $^{58}$Fe and $^{56}$Cr,
we have symmetric $3^+$ bases as well as MS $3^+$ ones.
The symmetric components contain three $D$-pairs at least,
while one of the MS bases consists only of two $D$-pairs.
It is of interest which component dominates
the lowest collective $3^+$ state.
According to the present calculation,
the lowest collective $3^+$ state is dominated
by the MS component in these nuclei,
with the excitation energy of 3.9--4.2~MeV,
as is presented in Table~\ref{tab:F-prob}.
As well as the excitation energy, 
the admixture of the symmetric component
is stable among the three nuclei (20--30\%).

The MS $1^+$ state is relatively easy
to observe\cite{ref:scissors2}.
It is worth discussing the position of the MS $3^+$ state
relative to the MS $1^+$ state.

In the IBM-2, the lowest MS $1^+$ and $3^+$ states
are approximately expressed as
\begin{eqnarray}
  |1_{\rm M}^+\rangle \propto [d_\pi^\dagger d_\nu^\dagger]^{(1)}
 \left|0_c^+ (N_\pi^{\rm B}-1, N_\nu^{\rm B}-1) \right\rangle, \\
  |3_{\rm M}^+\rangle \propto [d_\pi^\dagger d_\nu^\dagger]^{(3)}
 \left|0_c^+ (N_\pi^{\rm B}-1, N_\nu^{\rm B}-1) \right\rangle,
\end{eqnarray}
where the core state
$\left|0_c^+ (N_\pi^{\rm B}-1, N_\nu^{\rm B}-1) \right\rangle$
is an appropriately chosen totally-symmetric $0^+$ state
with $(N_\pi^{\rm B}-1)$ proton-bosons 
and $(N_\nu^{\rm B}-1)$ neutron-bosons.
If the nuclear shape does not strongly depend on the boson number,
$\left|0_c^+ (N_\pi^{\rm B}-1, N_\nu^{\rm B}-1) \right\rangle$ 
can be taken as the ground state with the corresponding boson numbers.
Neglecting the difference in coupling of the $d_\pi d_\nu$ part
to the $\left|0_c^+ (N_\pi^{\rm B}-1, N_\nu^{\rm B}-1) \right\rangle$
core, we reach an approximation for the relative energy as
\begin{eqnarray}
  E(3_{\rm M}^+) - E(1_{\rm M}^+) &=&
\langle 3_{\rm M}^+|H^{\rm B}|3_{\rm M}^+ \rangle
 - \langle 1_{\rm M}^+|H^{\rm B}|1_{\rm M}^+ \rangle \nonumber \\
 &\approx& \langle d_\pi d_\nu ; 3^+ |H^{\rm B}
 | d_\pi d_\nu ; 3^+ \rangle
- \langle d_\pi d_\nu ; 1^+ |H^{\rm B}| d_\pi d_\nu ; 3^+ \rangle ,
 \label{eq:3to1a}
\end{eqnarray}
with the boson hamiltonian denoted by $H^{\rm B}$.
Note that this relation holds exactly in the group-theoretical limits
$U(5)_{\pi+\nu}\otimes SU_F(2)$, $SU(3)_{\pi+\nu}\otimes SU_F(2)$
and $O(6)_{\pi+\nu}\otimes SU_F(2)$\cite{ref:IBM}.

We here assume the following IBM-2 hamiltonian,
\begin{equation}
 H^{\rm B} = \sum_{\rho=\pi,\nu} \epsilon_{d_\rho} \hat N_{d_\rho}
 - \kappa \hat Q_\pi \cdot \hat Q_\nu 
 + \sum_{J=1,2,3} \xi_J \hat M_J , \label{eq:HB}
\end{equation}
where
\begin{eqnarray}
  \hat Q_\rho &=& [d_\rho^\dagger s_\rho
    + s_\rho^\dagger \tilde d_\rho]^{(2)}
    + \chi_\rho [d_\rho^\dagger \tilde d_\rho]^{(2)} ,\\
  \hat M_J &=& [d_\pi^\dagger d_\nu^\dagger]^{(J)}
   \cdot [\tilde d_\nu \tilde d_\pi]^{(J)} ,~~\mbox{for $J=1,3$},\\
  \hat M_2 &=& {1\over 2} [d_\pi^\dagger s_\nu^\dagger
   - s_\pi^\dagger d_\nu^\dagger]^{(2)}
    \cdot [\tilde d_\pi s_\nu - s_\pi \tilde d_\nu]^{(2)} .
\end{eqnarray}
Though a more general IBM-2 hamiltonian has been derived 
in Ref.\cite{ref:no.1},
this form is usually sufficient to discuss the basic structure
of the low-lying states.
Substituting the above hamiltonian (\ref{eq:HB})
into Eq.~(\ref{eq:3to1a}), 
we obtain
\begin{equation} E(3_{\rm M}^+) - E(1_{\rm M}^+) \approx
 \kappa \chi_\pi \chi_\nu + (\xi_3 - \xi_1) . \end{equation}
The $\kappa$ parameter is positive in physical cases,
and is relatively insensitive to nuclide.
If the $(\xi_3-\xi_1)$ term is negligible,
the nucleus-dependence of $E(3_{\rm M}^+)$ 
relative to $E(1_{\rm M}^+)$
is essentially governed by the $\chi_\pi \chi_\nu$ value.

The large positive $\chi_\pi \chi_\nu$ value in $^{56}$Fe,
which has been shown in Table~\Roman{ref2} of Ref.\cite{ref:no.1},
makes the MS $3^+$ state substantially higher
than the MS $1^+$ state.
The difference between $\xi_1$ and $\xi_3$ in this nucleus
is not large enough to change this situation.
On the other hand, we have $\chi_\pi \chi_\nu \approx 0$
in $^{54}$Cr, $^{58}$Fe and $^{56}$Cr.
It should also be noticed that $\xi_1 \approx \xi_3$ in these nuclei.
Thereby comparable excitation energy is expected
between the MS $1^+$ and $3^+$ states.
Owing to admixture of the symmetric component,
the MS-dominant $3^+$ state appears
even lower than the $1^+$ state.

\section{Distributions of mixed-symmetry states}
\label{sec:dist}

We have explicitly constructed
the wavefunctions of the collective fermion states
via the H$^2$CM.
This enables us to compare these states 
with the shell-model eigenstates,
by calculating overlaps between their wavefunctions.
In this section we shall investigate
how the lower-lying collective states dominated by the MS components
distribute over the shell-model eigenstates.

The overlaps of the MS $2^+$ state of $^{56}$Fe
with the shell-model eigenstates in $E_x<8$~MeV
are shown in Fig.~\ref{fig:Fe56-dist}(a).
A certain fragmentation of this $2^+$ component occurs,
because of the coupling to non-collective $2^+$ states 
in its vicinity.
It is found, however, that the main fraction is shared
by the $2_2^+$ and $2_4^+$ states with more than $70\%$ probability
in total.
This is consistent with the previous report in Ref.\cite{ref:NOS91}.
As is viewed in Table~\ref{tab:prob-sum},
the summed probability of the MS $2^+$ component
in the shell-model eigenstates up to 5~MeV reaches as much as 80\%.
The fact that the collective-state amplitudes are concentrated
in a relatively small energy range
shows that we can regard the MS state as a basic building block.
We can interpret this state as a doorway-type state.

Similarly, 
the doorway nature of the MS $1^+$ state is clarified
in the distribution over the shell-model eigenstates,
as shown in Fig.~\ref{fig:Fe56-dist}(b).
The $1^+$ component is mainly shared
by the shell-model $1_2^+$ and $1_3^+$ states,
both of which exist around 3.5~MeV, with 87\% probability in total.

The MS $1^+$ component can be searched
by using the $M1$ transition to the ground state.
Whereas the $M1$ transition is forbidden
for a purely spherical ground-state,
even a small deformation makes it measurable.
The following shell-model $M1$ operator is adopted
as in Refs.\cite{ref:NOS91,ref:no.1},
\begin{equation}
T(M1) = \sqrt{{3\over{4\pi}}} \sum_{\rho=\pi,\nu} \left\{
g_{l,\rho}^{\rm eff} \sum_{i\in\rho} l_i
 + g_{s,\rho}^{\rm eff} \sum_{i\in\rho} s_i \right\} ,
\label{eq:M1op} \end{equation}
together with $g_{l,\pi}^{\rm eff}=g_{l,\pi}^{\rm free}=1.0[\mu_N]$,
$g_{l,\nu}^{\rm eff}=g_{l,\nu}^{\rm free}=0.0[\mu_N]$,
$g_{s,\pi}^{\rm eff}=0.5 g_{s,\pi}^{\rm free}$
and $g_{s,\nu}^{\rm eff}=0.5 g_{s,\nu}^{\rm free}$.
By employing this $M1$ operator with the shell-model wavefunctions,
the $M1$ transition strengths within the $k\leq 2$ space
are calculated and depicted in Fig.~\ref{fig:Fe56-tr},
with $k$ denoting the number of nucleons excited out of $0f_{7/2}$.
It should be recalled that the $M1$ transition
between totally symmetric states are forbidden
within the IBM-2 framework\cite{ref:Scholten}.
On the other side, the $M1$ transition
between the ground state and the state with large MS component
is expected to be measurable.
Although the $1^+$ states consisting of the $k=3$ configuration
carry some $M1$ strengths,
their contribution will be important only in the higher energy region,
as will be discussed further in Sect.~\ref{sec:M1-dist}.
The above spin-quenching factor has been fitted
to the $B(M1)$ values among lowest-lying states 
of $^{56}$Fe\cite{ref:NOS91}.
This factor is considerably smaller
than the one predicted from microscopic standpoints\cite{ref:Towner},
and may be ascribed to the influence of higher $k$ configurations.
It is noticed that a relatively strong influence
of higher $k$ configurations has been suggested for $^{56}$Fe
on the basis of the electromagnetic properties\cite{ref:NSO94}.
It is found that the $B(M1)$ distribution over low-lying states
resembles the distribution of the MS component
shown in Fig.~\ref{fig:Fe56-dist}(b).
Both of the $1_2^+$ and $1_3^+$ states
have relatively large $B(M1)$ values to the ground state;
$0.04[\mu_N^2]$ from $1_2^+$ and $0.10[\mu_N^2]$ from $1_3^+$.
The concentration of the MS $1^+$ component
around $E_x\simeq 3.5$~MeV
seems to be consistent with the recent $(e,e')$ 
and $(p,p')$ experiments
reported in Ref.\cite{ref:Fe56MS1}.

The distribution of the second $2^+$ state 
in the collective space of $^{54}$Cr
over the shell-model eigenstates
is depicted in Fig.~\ref{fig:Cr54-dist}(a).
The shell-model $2_3^+$ state absorbs
about ${2\over 3}$ of the MS component,
and more than $90\%$ is exhausted by the states below 5~MeV,
as shown in Table~\ref{tab:prob-sum}.
The MS-dominant $2^+$ state also remains a basic mode
through the doorway interpretation.

The lowest collective $1^+$ and $3^+$ components of $^{54}$Cr
are confirmed to have similar nature,
as presented in Fig.~\ref{fig:Cr54-dist}(b) and (c).
The shell-model $1_3^+$ and $3_1^+$ states
have large fractions of the MS states.
Although both these states have not been observed,
their MS-dominance is probably seen experimentally
by the $M1$ or $M3$ excitation.
As well as the $M1$ transition,
the $M3$ transition may be sizable
between the ground state and the MS state.
According to the shell-model calculation,
we predict $B(M1; 1_3^+ \rightarrow 0_1^+) \simeq 0.23 [\mu_N^2]$
and $B(M3; 3_1^+ \rightarrow 0_1^+) \simeq 460 [\mu_N^2 {\rm fm}^4]$,
by assuming the bare $M3$ operator
\begin{equation} T(M3) = {\sqrt{21}\over 2} \sum_{\rho=\pi,\nu}
 \left\{ g_{l,\rho}^{\rm free} \sum_{i\in\rho}
  r_i^2 [Y^{(2)}(\hat{\bf r}_i) l_i]^{(3)}
 + 2g_{s,\rho}^{\rm free} \sum_{i\in\rho}
  r_i^2 [Y^{(2)}(\hat{\bf r}_i) s_i]^{(3)} \right\} ,
\label{eq:M3op} \end{equation}
and the harmonic-oscillator single-particle wavefunctions
with $b=1.956$fm, as in Ref.\cite{ref:no.1}.
The distribution of the low-lying $M1$ and $M3$ strengths 
is presented in Fig.~\ref{fig:Cr54-tr},
for the convenience of experimental search.
The $B(M3)$ value from $3_1^+$ to the ground state
is even larger than the Weisskopf unit,
which is $337 [\mu_N^2 {\rm fm}^4]$ at $A=54$.
Moreover, the $M3$ transition from $3_1^+$ is notably stronger
than those from neighboring $3^+$ states,
which may help us to identify the $3_1^+$ state 
through a scattering experiment.
We have a few strong $M3$ strengths below 6~MeV,
for states which have small fraction of the MS component.
This is a notable contrast to the $M1$ transition,
which has a clear correlation to the distribution of the MS component.
This happens mainly because the spin term in Eq.~(\ref{eq:M3op})
has larger effects than in the $M1$ case.

Figures~\ref{fig:Fe58-dist} and \ref{fig:Cr56-dist} show
the distributions of the collective states dominated
by the MS components
over the shell-model eigenstates, up to around 4~MeV,
for the $N=32$ nuclei.
Summed strengths are presented in Table~\ref{tab:prob-sum}.
Because of a numerical difficulty caused
by the greater model space\cite{ref:no.1},
the energy range in which the shell-model eigenstates are searched
is restricted to a smaller region than in the $N=30$ nuclei.
We mention a few points at this stage.
There is a certain fragmentation
for any of the collective MS-dominant states
in the renormalized $SD$ space.
The fragmentation appears somewhat stronger in these $N=32$ nuclei
than in the $N=30$ nuclei.
Nevertheless, an appreciable portion of the collective component
is already found in the low energy region,
as is shown in Table~\ref{tab:prob-sum}.
We remark that the investigated energy range 
does not sufficiently exceed
the energies of the lowest MS-dominant states
in the renormalized $SD$ space.
This could be the reason why the probabilities of the MS component
are so small for the $N=32$ nuclei,
in comparison with the $N=30$ nuclei.

As stated in Sect.~\ref{sec:intro},
the fragmentation of the MS component over a small energy range
has been observed for the scissors $1^+$ states 
of heavier rotational nuclei.
The situation looks quite similar to the current Cr-Fe case,
suggesting the following global nature of lower-lying MS states.
It is inevitable for this type of collective components
to fragment more or less,
because there exist a certain number 
of non-collective degrees-of-freedom
in the same energy region.
While the coupling to the non-collective degrees-of-freedom
leads to a certain fragmentation,
the coupling is not so strong
as for the MS states to lose their identity.
The main fraction of the MS components
remains in the vicinity of the original energy,
and it is shared primarily by a few eigenstates.
It is thus reasonable to regard the MS state
as a doorway state.

\section{$B(M1)$ distribution in $^{56}$F\lowercase{e}
and $^{54}$C\lowercase{r}}
\label{sec:M1-dist}

The $M1$ transition seems to be a good probe
in investigating the MS $1^+$ component. 
As has been shown in Sect.~\ref{sec:dist},
the low-lying $M1$ transition strengths
correspond well to the MS $1^+$ component.
In this section, we research the $B(M1)$ distribution
covering a wider energy range,
which will be useful for future experiments.
We shall restrict ourselves to $^{56}$Fe and $^{54}$Cr,
in order to avoid computational difficulties 
in $^{58}$Fe and $^{56}$Cr.

\subsection{Summed strengths and central energies}
\label{subsec:sum}

As in the Gamow-Teller (GT) transition\cite{ref:NS96},
a certain part of the $M1$ strengths is carried
by the $1^+$ states with the $k=3$ configuration,
even if the ground state is described
only by the $k\leq 2$ configurations.
Therefore, when we study the $B(M1)$ distribution
including a relatively high energy region
with the $k\leq 2$ ground-state wavefunction,
we should not discard the $k=3$ configuration for the $1^+$ states.
For this reason,
in the following we treat the $1^+$ states in the $k\leq 3$ space,
keeping the ground state in the $k\leq 2$ space.
In fact, there should be a certain admixture 
of the $k=3$ configuration
in the low-lying $1^+$ states,
although their influence will be small and may be taken into account
by adjusting the single-particle parameters.
Though the $B(M1)$ values among low-lying states
have been reproduced by the $M1$ operator of Eq.~(\ref{eq:M1op})
within the $k\leq 2$ space,
the single-particle parameters adopted there
reflect the influence of the $k=3$ configuration to some extent.
There is no reason to apply the same parameters to the calculation
involving the $k=3$ configuration explicitly.
We here use, for convenience, the $M1$ operator
with the single-particle parameters evaluated 
by Towner\cite{ref:Towner}
from microscopic viewpoints.

Table~\ref{tab:M1-sum} exhibits, for $^{56}$Fe and $^{54}$Cr,
the summed $B(M1)$ values
from all the possible $1^+$ states to the ground state.
The summed $M1$ strength for each isospin component is shown,
as well as the total strength.
Note that, if the isospin of the ground state is denoted by $T_0$
({\it i.e.}, $T_0=2$ for $^{56}$Fe and $T_0=3$ for $^{54}$Cr),
$T=T_0$ and $T_0+1$ are possible for the $1^+$ states.
The summed $B(M1)$ values are actually calculated as follows:
we first generate the following state,
which exhaust the non-energy-weighted sum,
\begin{equation} |1^+_{\rm sum}\rangle \equiv
 {\cal N}~T(M1) |0^+_1\rangle , \end{equation}
where ${\cal N}$ stands for a normalization constant.
The total $B(M1)$ value is equal to ${\cal N}^{-2}$.
We next carry out the isospin projection for $|1^+_{\rm sum}\rangle$,
and the probability of each isospin component 
in $|1^+_{\rm sum}\rangle$
gives the ratio of the $M1$ strengths
between the isospin-conserving ($T=T_0$)
and isospin-raising ($T=T_0+1$) components.
If one wishes to obtain the $M1$ excitation strengths
from the ground state,
the shown $B(M1)$ values should be multiplied by a factor of three.
It is noted that the isospin decomposition of the $M1$ strength
is important in some cases;
for example, it yields significant information
on the double-$\beta$-decay matrix-element\cite{ref:NSM96},
via the close relation of $M1$ to the GT transition.

Table~\ref{tab:M1-sum} also shows the central energies
of the $M1$ strengths,
\begin{equation} \bar{E_x} \equiv
{{\sum_i E_x(1^+_i) B(M1;1^+_i\rightarrow 0^+_1)}\over
{\sum_i B(M1;1^+_i\rightarrow 0^+_1)}} . \end{equation}
At first glance, this appears to be obtained by
$\langle 1^+_{\rm sum}|P_T HP_T|1^+_{\rm sum}\rangle
/ \langle 1^+_{\rm sum}|P_T|1^+_{\rm sum}\rangle$,
where $H$ indicates the shell-model hamiltonian 
in the $k\leq 3$ space
and $P_T$ denotes the isospin projector.
However, the following corrections are necessary.
There is a certain difference in the ground-state energy
between the $k\leq 2$ and $k\leq 3$ spaces.
Although $\bar{E_x}$ should be measured
from the $k\leq 3$ ground-state energy,
it is not easy to compute the ground-state energy
in the $k\leq 3$ space because of its large dimensionality
(about $2\times 10^6$ in the $M$-scheme). 
It is remarked that,
as far as the energy intervals among low-lying $1^+$ states are 
concerned,
the $k\leq 3$ space has been confirmed
to yield almost the same values as the $k\leq 2$ space.
To circumvent the time-consuming computation,
we shift the $T=T_0$ $1^+$ levels
so that $E_x(1^+_1)$ in the $k\leq 3$ space
becomes equal to that obtained in the $k\leq 2$ space\cite{ref:NSO94}.
Regarding the $T=T_0+1$ case, we adjust the $T=T_0+1$ levels,
by using the data of the neighboring nuclei\cite{ref:IsoTab}.
For $^{56}$Fe, the energy of the lowest $1^+$ state with $T=3$
is extracted from $E_x(1^+_1)$ of $^{56}$Mn
and the mass difference between $^{56}$Fe and $^{56}$Mn.
For the Coulomb-energy difference,
we apply the formula derived from the uniform charge-distribution,
\begin{equation} \Delta E_{\rm C} =
 {{e^2}\over{4\pi}} {3\over{5 r_{\rm C} A^{1/3}}} [Z^2-(Z-1)^2]
 \simeq 0.696 (2Z-1) A^{-1/3}~{\rm MeV}~~(r_{\rm C}=1.24{\rm fm}).
\end{equation}
The energy of the $T=4$ $1^+$ component of $^{54}$Cr
is corrected in a similar manner, by using the $^{54}$V data.
Although the $1^+_1$ level has not been confirmed 
in $^{54}$V\cite{ref:IsoTab},
we assume that the second excited state is $1^+_1$
according to systematics of $N=31$ nuclei.
The estimate of the Coulomb-energy difference 
might not be very precise.
However, the presented $\bar{E_x}$ values
could be a guidance to future experiments.

\subsection{Mixed-symmetry component}
\label{subsec:M1-MS}

The $B(M1)$ distribution is calculated for each isospin component
via Whitehead's moment method\cite{ref:Wh80},
starting from $P_T|1^+_{\rm sum}\rangle$ with 45 iterations.
The results are given in Figs.~\ref{fig:Fe56-fullM1}
and \ref{fig:Cr54-fullM1},
where the $B(M1)$ values to the ground state
are displayed by bins of 1~MeV.
The spin-excitation strengths
are calculated by setting $g_{l,\rho}=0$ $(\rho=\pi,\nu)$
in the $M1$ operator,
and are also displayed in Figs.~\ref{fig:Fe56-fullM1}
and \ref{fig:Cr54-fullM1}.

For $^{56}$Fe presented in Fig.~\ref{fig:Fe56-fullM1},
the energies of the lowest three $1^+$ levels are convergent
in the iteration of Whitehead's moment method.
As well as the energy intervals among these levels,
the $B(M1)$ values are close to those given in Sect.~\ref{sec:dist};
and $B(M1;1^+_3\rightarrow 0^+_1)=0.11\mu_N^2$.
Thus, for the $M1$ strengths from these low-lying states,
the influence of the difference in size of the model space
is absorbed into the single-particle parameters.

In Fig.~\ref{fig:Fe56-fullM1},
we see a low peak at 3--4~MeV for the $T=T_0=2$ part,
well below the main peak at around 8~MeV.
The excitation regarding the orbital angular-momentum
highly dominates this peak.
The $1^+_2$ and $1^+_3$ states form this low-energy peak.
As it has been shown in Sect.~\ref{sec:dist},
the $1^+_2$ and $1^+_3$ share the main fraction of the MS component.
Notice that the MS component has $T=T_0=2$,
and has been described by the $k\leq 2$ configurations 
in Sect.~\ref{sec:dist},
which should dominate the low-lying states.
Therefore, the distribution of the MS $1^+$ component 
shown in Sect.~\ref{sec:dist}
is connected to the low-energy part of the entire $M1$ distribution.

For $^{54}$Cr, the energies of the low-lying $1^+$ levels
are not sufficiently convergent with 45 iterations.
Still, the energy intervals among several lowest-lying components
appear to be similar to those obtained in the $k\leq 2$ space.
We also notice a peak around 4~MeV
separated from the main peak (see Fig.~\ref{fig:Cr54-fullM1}),
thus indicating that the MS component is observable
via the low-energy $M1$ peak.
This peak is predominantly constituted
by the excitation of the orbital angular-momentum,
as in $^{56}$Fe.

\subsection{Spin excitation}
\label{subsec:spin}

We next turn to the main peaks of the $B(M1)$ distribution.
The main peak for the $T=T_0$ strengths
(see Figs.~\ref{fig:Fe56-fullM1} and \ref{fig:Cr54-fullM1})
is dominated by the single-particle excitation
from the $j=l+1/2$ orbit to its spin-orbit partner $j'=l-1/2$.
Such an excitation is often called spin excitation,
because it is mainly contributed by the nucleon-spin operator,
as shown below.
The high-energy tail of the peak is low
but damps very gradually.
The $B(M1)$ value up to 12~MeV amounts to 93\% (94\%)
of the whole sum of $T=T_0$ strengths, for $^{56}$Fe ($^{54}$Cr).
Therefore caution will be necessary
in looking at experimental data on the summed $B(M1)$ values:
it is hard to extract the isospin-conserving $M1$ strength
by experiments with better precision than 10\%.

For $^{56}$Fe, the $M1$ distribution below 15~MeV
has been investigated experimentally\cite{ref:scissors2}.
The pattern of the distribution is reproduced well
by the present calculation.

It is found in Fig.~\ref{fig:Fe56-fullM1} that,
around the main peak, the $M1$ strengths due only to the spin terms
are substantially greater than the $B(M1)$ values
containing all contributions.
Namely, although the spin contribution is dominant in this peak,
there exists a certain destructive contribution
of the orbital angular-momentum.
This is explained in a simple way as follows.
Let us consider the single-particle matrix-element
$\langle j'=l-1/2||T(M1)||j=l+1/2\rangle$.
Since $\langle j'=l-1/2||j||j=l+1/2\rangle=0$,
we have $\langle j'=l-1/2||l||j=l+1/2\rangle=
-\langle j'=l-1/2||s||j=l+1/2\rangle$.
Neglecting the $[Y^{(2)}s]$-term in the $M1$ operator,
we obtain $\langle j'=l-1/2||T(M1)||j=l+1/2\rangle\propto
(g_s-g_l)$\cite{ref:BM1}.
Because $|g_s|$ is appreciably larger than $|g_l|$,
the spin contribution is dominant in this excitation,
as has been stated above.
On the other hand, the orbital contribution should be present,
as far as the $M1$ transition involves proton excitations.
Therefore, except for the proton $LS$-closed nuclei,
the spin excitation should contain a destructive contribution
from the orbital angular-momentum.
We thus need caution
in extracting the $B(M1)$ values from $(p,p')$ experiments,
which hardly excite the orbital motion
owing to the locality of the $NN$ interaction:
$(p,p')$ tends to overestimate the $B(M1)$ values
around the peak region.
For $^{54}$Cr (see Fig.~\ref{fig:Cr54-fullM1}),
the destructive orbital contribution to the spin excitation
is confirmed by summing the $M1$ strengths in the 7--10~MeV region,
though it is not apparent for individual bins
because of the poor convergence.

The main fraction of the $T=T_0+1$ strengths,
which are also shown in Fig.~\ref{fig:Fe56-fullM1} for $^{56}$Fe
and Fig.~\ref{fig:Cr54-fullM1} for $^{54}$Cr,
seems to be distinct in energy 
from the main body of the $T=T_0$ strengths,
besides the reliability of the estimated excitation energy.
The peak appears to be high enough
as not to be hidden in the $T=T_0$ component.
However, the tail of the $T=T_0$ component amounts to about 10\%
for each bin relative to the $T=T_0+1$ peak height.
Even if the peak can be observed clearly,
it will not be easy to obtain the isospin-raising strength
from $M1$ excitation experiments with a high precision.
While the spin degrees-of-freedom give the main contribution
to the $T=T_0+1$ $M1$ strengths,
the orbital angular-momentum contributes destructively,
by the same mechanism as in the $T=T_0$ case.

\section{Summary and discussion}
\label{sec:summary}

By applying the $H^n$-cooling method (H$^n$CM) with $n=2$,
the IBM-2 picture is extracted
from the large-scale shell-model results
for $^{56}$Fe, $^{54}$Cr, $^{58}$Fe and $^{56}$Cr.
The lower-lying MS states are investigated
with a particular interest:
the distributions of these components are seen
in terms of the overlaps between the wavefunctions
in the renormalized $SD$ space
and the shell-model eigenfunctions.
In $^{56}$Fe and $^{54}$Cr,
the MS states are shown to be basic modes,
while weak fragmentation is inevitable 
because of the high level density.
Therefore, the doorway interpretation is appropriate 
for the MS states.
This consequence is consistent
with the behavior of the scissors $1^+$ state observed 
in heavier nuclei.
Despite the smaller energy range of investigation,
there is no contradiction either in $^{58}$Fe nor $^{56}$Cr
with the doorway nature of the lower-lying MS states.

In addition to Ref.\cite{ref:NOS91},
Halse studied quadrupole collective modes including the MS ones
in the Cr-Fe nuclei\cite{ref:Halse}.
His work has been based on 
a more severely truncated shell-model calculation
assuming the $^{56}$Ni inert core\cite{ref:HO},
which does not reproduce the energy levels beyond 3~MeV precisely.
The predicted energies of the MS-dominant states
are not so different from the present one,
though the fragmentation of those components was out of scope 
in that work.

We show a clear evidence that the lowest MS-dominant state
has $J^P=2^+$ for these nuclei.
There have been several theoretical and experimental suggestions,
in this mass region\cite{ref:mixsym2,ref:Halse}
as well as in heavier-mass region\cite{ref:N=84},
that the lowest MS state has $J^P=2^+$ for some nuclei.
This expectation is confirmed by the present realistic study.
The indication of the MS $3^+$ state is also obtained.
It is almost the first realistic calculation on the MS $3^+$ state,
except the study based on the $M3$ transition
in lighter {\em pf}-shell region\cite{ref:Zamick}.

The low-lying $M1$ strengths to the ground state are calculated
for $^{56}$Fe and $^{54}$Cr.
The $M1$-strength distribution resembles the distribution
of the MS $1^+$ component.
It is hopeful to search the MS $1^+$ component experimentally.
The $M3$ strength is also calculated for $^{54}$Cr.
Despite a certain additional contribution
of non-collective degrees-of-freedom,
the MS $3^+$ component may be detectable
by scattering experiments.
The $B(M1)$ distribution in a wider energy range,
as well as the summed $B(M1)$ values, are also presented,
for the isospin-conserving and isospin-raising components,
respectively.
Unlike the RPA approach, we can treat the isospin correctly.
The MS $1^+$ component forms a peak in the low-energy region,
and is well separated from the main peak of the spin excitation.
Concerning the summed $M1$ strength,
it is found that the tail of the $B(M1)$ distribution
may contribute by about 10\%.
The isospin-raising component seems to have a distinct peak
from the isospin-conserving one.
The destructive contribution of the orbital angular-momentum
is confirmed both for the isospin-conserving and isospin-raising
spin excitations.

\acknowledgments
The authors are grateful to Prof. A. Gelberg
for careful reading the manuscript.

\clearpage
\begin{table}
\centering
\caption{$F$-spin probabilities (\%) of the states
in the renormalized $SD$ space via the H$^2$CM.
\label{tab:F-prob}}
\begin{tabular}{ccrrr}
    Nucleus & $J^P$ & \multicolumn{1}{c}{$E_x$(MeV)} &
 \multicolumn{1}{c}{\hspace{3cm} $F=F_{\rm max}$} &
 \multicolumn{1}{c}{$F=F_{\rm max}-1$} \\
   \hline
   $^{56}$Fe & $0^+$ & $0.000~~$ & $100.0~~$ & $0.0~~$ \\
   & $1^+$ & $3.746~~$ & $0.0~~$ & $100.0~~$ \\
   & $2^+$ & $1.088~~$ & $99.9~~$ & $0.1~~$ \\
   & $2^+$ & $3.568~~$ & $0.1~~$ & $99.9~~$ \\
   & $3^+$ & $4.835~~$ & $0.0~~$ & $100.0~~$ \\
   \hline
   $^{54}$Cr & $0^+$ & $0.000~~$ & $91.0~~$ & $9.0~~$ \\
   & $1^+$ & $4.378~~$ & $0.0~~$ & $100.0~~$ \\
   & $2^+$ & $1.034~~$ & $92.1~~$ & $7.9~~$ \\
   & $2^+$ & $3.171~~$ & $44.7~~$ & $55.3~~$ \\
   & $3^+$ & $4.148~~$ & $20.1~~$ & $79.9~~$ \\
   \hline
   $^{58}$Fe & $0^+$ & $0.000~~$ & $95.2~~$ & $4.8~~$ \\
   & $1^+$ & $4.707~~$ & $0.0~~$ & $100.0~~$ \\
   & $2^+$ & $1.171~~$ & $91.4~~$ & $8.6~~$ \\
   & $2^+$ & $2.689~~$ & $69.0~~$ & $31.0~~$ \\
   & $2^+$ & $3.697~~$ & $37.7~~$ & $62.3~~$ \\
   & $3^+$ & $3.922~~$ & $25.2~~$ & $74.8~~$ \\
   \hline
   $^{56}$Cr & $0^+$ & $0.000~~$ & $93.2~~$ & $3.4~~$ \\
   & $1^+$ & $4.786~~$ & $0.0~~$ & $100.0~~$ \\
   & $2^+$ & $1.301~~$ & $92.6~~$ & $4.7~~$ \\
   & $2^+$ & $2.730~~$ & $81.8~~$ & $15.3~~$ \\
   & $2^+$ & $3.462~~$ & $13.8~~$ & $83.6~~$ \\
   & $3^+$ & $4.108~~$ & $26.4~~$ & $72.3~~$ \\
\end{tabular}
\end{table}

\begin{table}
\centering
\caption{Summed probabilities of the MS-dominant components
over the shell-model eigenstates.
The energy range for the summation is also shown.
\label{tab:prob-sum}}
\begin{tabular}{cccr}
    Nucleus & $J^P$ & Energy range &
 \multicolumn{1}{c}{\hspace{5cm} Prob. (\%)} \\
   \hline
   $^{56}$Fe & $1^+$ & $E_x<5$~MeV & $87~~$ \\
   & $2^+$ & $E_x<5$~MeV & $79~~$ \\
   \hline
   $^{54}$Cr & $1^+$ & $E_x<5$~MeV & $83~~$ \\
   & $2^+$ & $E_x<5$~MeV & $94~~$ \\
   & $3^+$ & $E_x<5$~MeV & $88~~$ \\
   \hline
   $^{58}$Fe & $1^+$ & $E_x<4.5$~MeV & $63~~$ \\
   & $2^+$ & $E_x<3.8$~MeV & $68~~$ \\
   & $3^+$ & $E_x<4$~MeV & $41~~$ \\
   \hline
   $^{56}$Cr & $1^+$ & $E_x<4.5$~MeV & $46~~$ \\
   & $2^+$ & $E_x<3.9$~MeV & $78~~$ \\
   & $3^+$ & $E_x<4.1$~MeV & $57~~$ \\
\end{tabular}
\end{table}

\begin{table}
\centering
\caption{Summed $B(M1)$ values and their central energies.
\label{tab:M1-sum}}
\begin{tabular}{ccrr}
    Nucleus && \multicolumn{1}{c}{$\sum B(M1)$~($\mu_N^2$)}
   & \multicolumn{1}{c}{$\bar{E_x}$~(MeV)} \\
   \hline
   $^{56}$Fe & $T=2$ & 4.22~~ & 9.1~~ \\
   & $T=3$ & 0.88~~ & 18.2~~ \\
   & total & 5.10~~ & 10.7~~ \\
   \hline
   $^{54}$Cr & $T=3$ & 4.20~~ & 8.6~~ \\
   & $T=4$ & 0.24~~ & 29.0~~ \\
   & total & 4.44~~ & 9.7~~ \\
\end{tabular}
\end{table}

\clearpage
\section*{Figure Captions}
\begin{list}{Fig.\thefig:\hfill}{\usecounter{fig}}
\item
Distribution of the MS-dominant components 
over the shell-model eigenstates in $^{56}$Fe:
(a) the second $2^+$ state and (b) the $1^+$ state,
obtained in the renormalized $SD$ space via the H$^2$CM.
The vertical axis shows squares of the overlaps.
\label{fig:Fe56-dist}
\item
Distribution of the MS-dominant components
over the shell-model eigenstates in $^{54}$Cr:
(a) the second $2^+$ state, (b) the first $1^+$ state
and (c) the first $3^+$ state,
obtained in the renormalized $SD$ space.
\label{fig:Cr54-dist}
\item
$B(M1)$ distribution from $1^+$ states to the ground state
in $^{56}$Fe, within the $k\leq 2$ space.
\label{fig:Fe56-tr}
\item
(a) $B(M1)$ and (b) $B(M3)$ distribution to the ground state
in $^{54}$Cr, within the $k\leq 2$ space.
\label{fig:Cr54-tr}
\item
Distribution of the MS-dominant components 
over the shell-model eigenstates in $^{58}$Fe:
(a) the third $2^+$ state, (b) the first $1^+$ state
and (c) the first $3^+$ state,
obtained in the renormalized $SD$ space.
\label{fig:Fe58-dist}
\item
Distribution of the MS-dominant components 
over the shell-model eigenstates in $^{56}$Cr:
(a) the third $2^+$ state, (b) the first $1^+$ state
and (c) the first $3^+$ state,
obtained in the renormalized $SD$ space.
\label{fig:Cr56-dist}
\item
$B(M1)$ distribution to the ground state in $^{56}$Fe,
for $1^+$ states with $T=2$ and $T=3$.
The values without orbital contribution are
indicated by the short bars.
\label{fig:Fe56-fullM1}
\item
$B(M1)$ distribution to the ground state in $^{54}$Cr,
for $1^+$ states with $T=3$ and $T=4$.
The values without orbital contribution are
indicated by the short bars.
\label{fig:Cr54-fullM1}
\end{list}

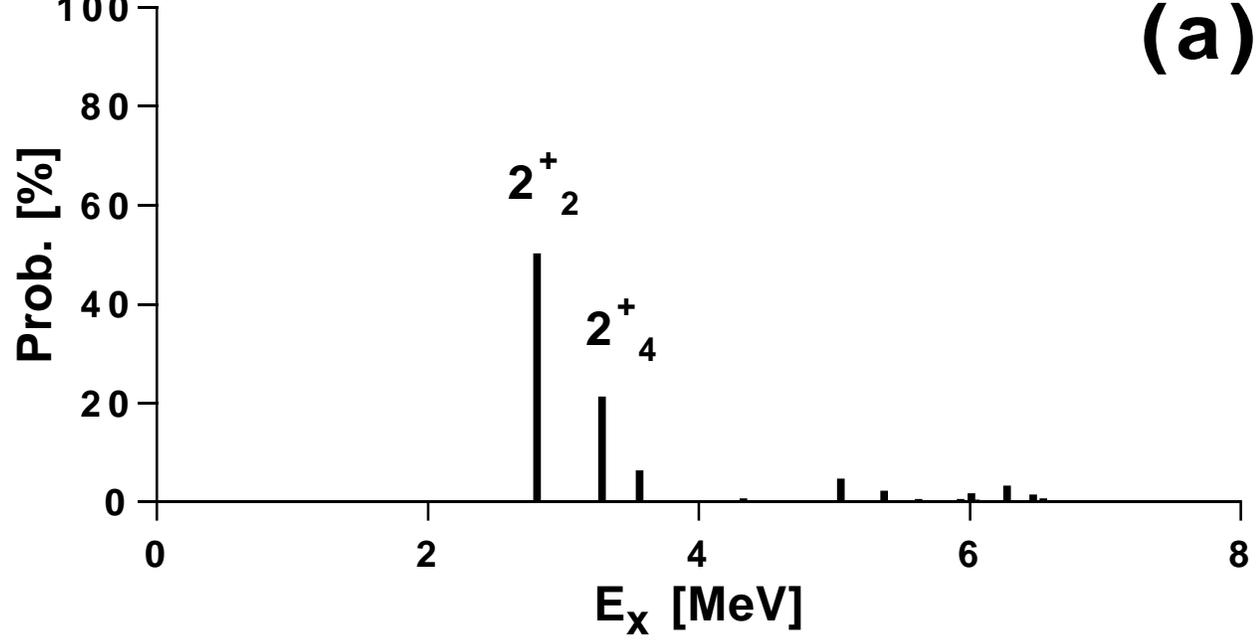
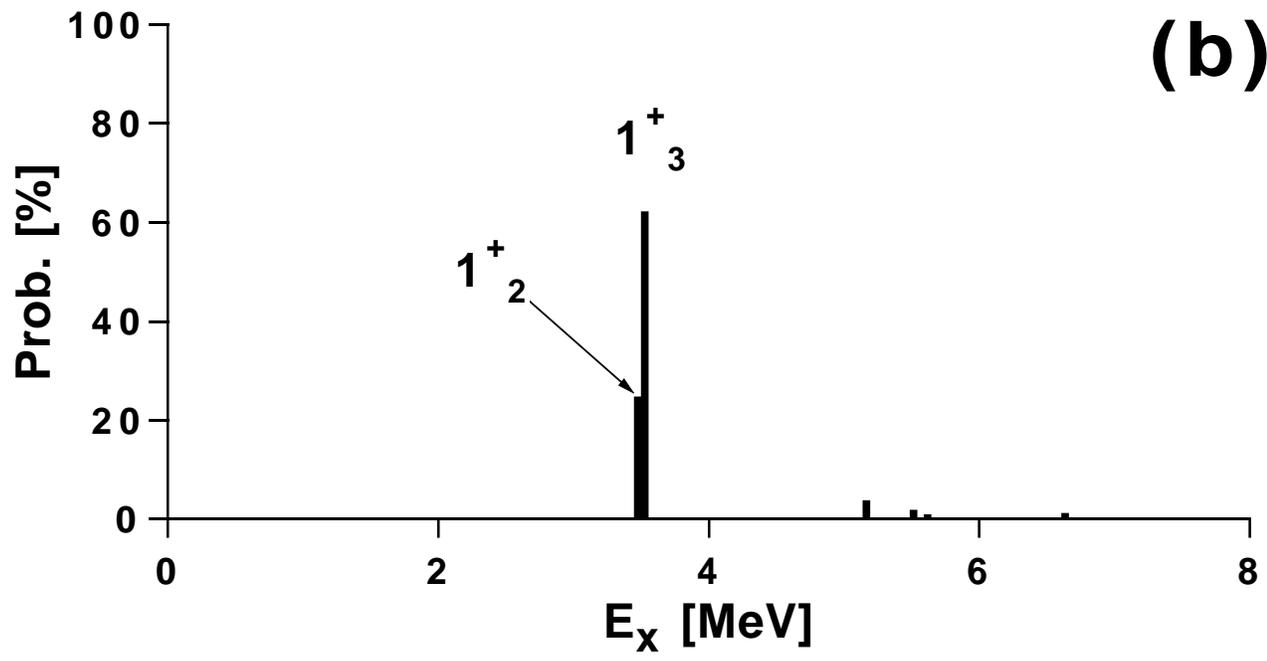
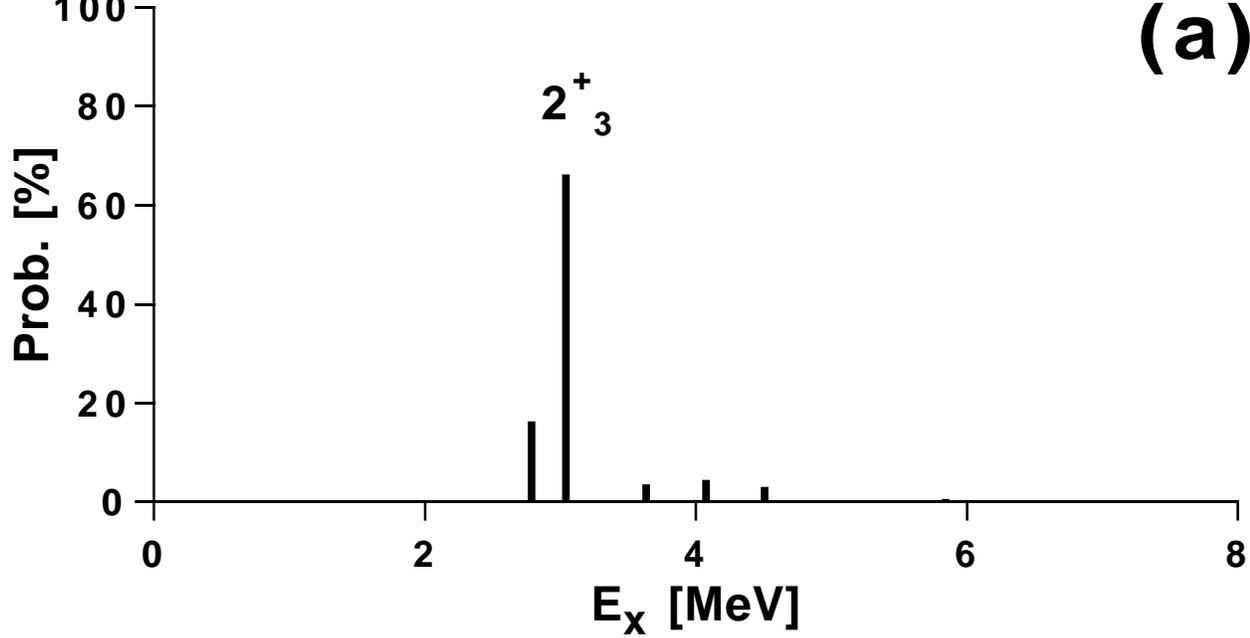
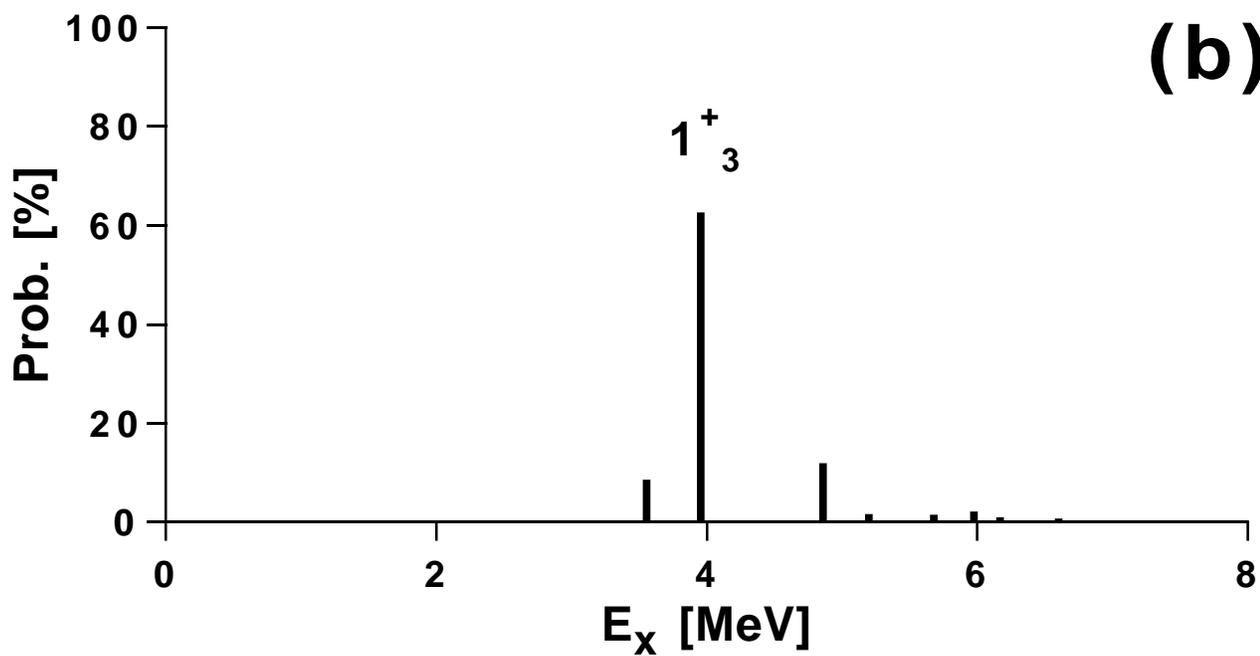
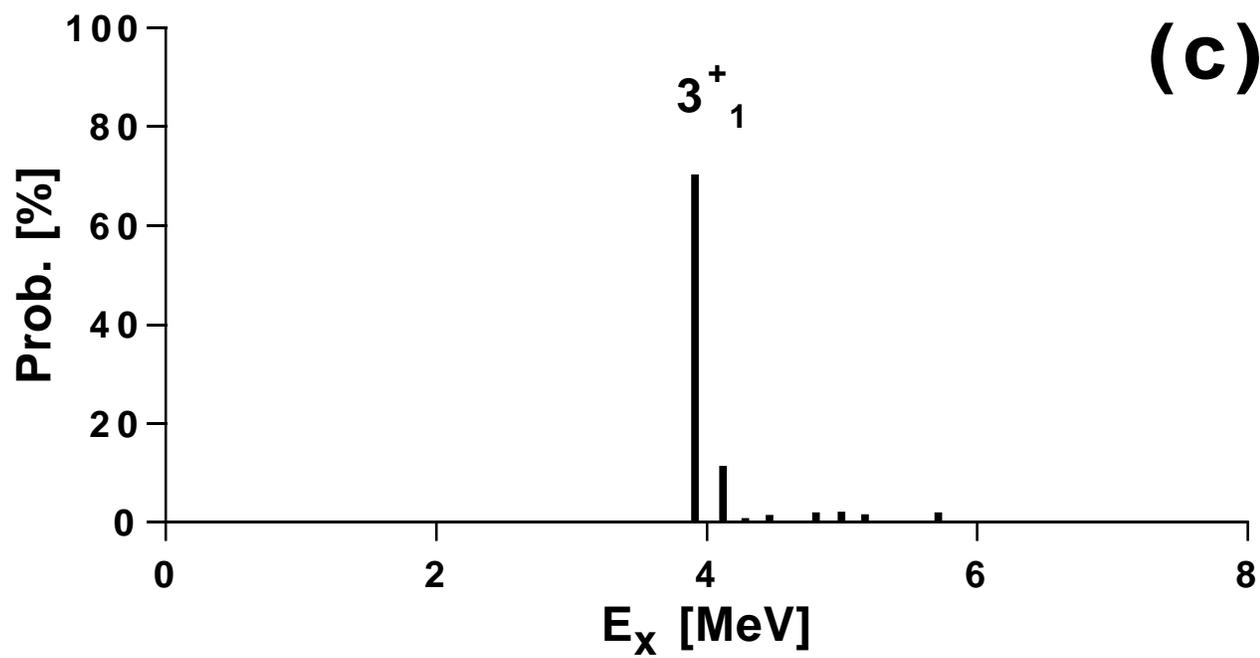
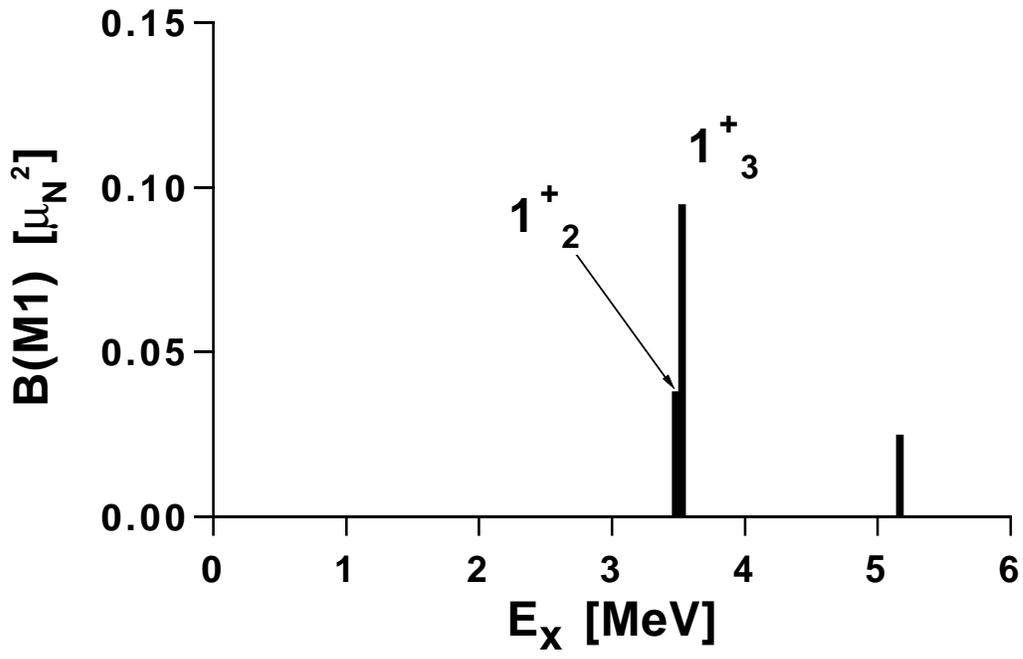
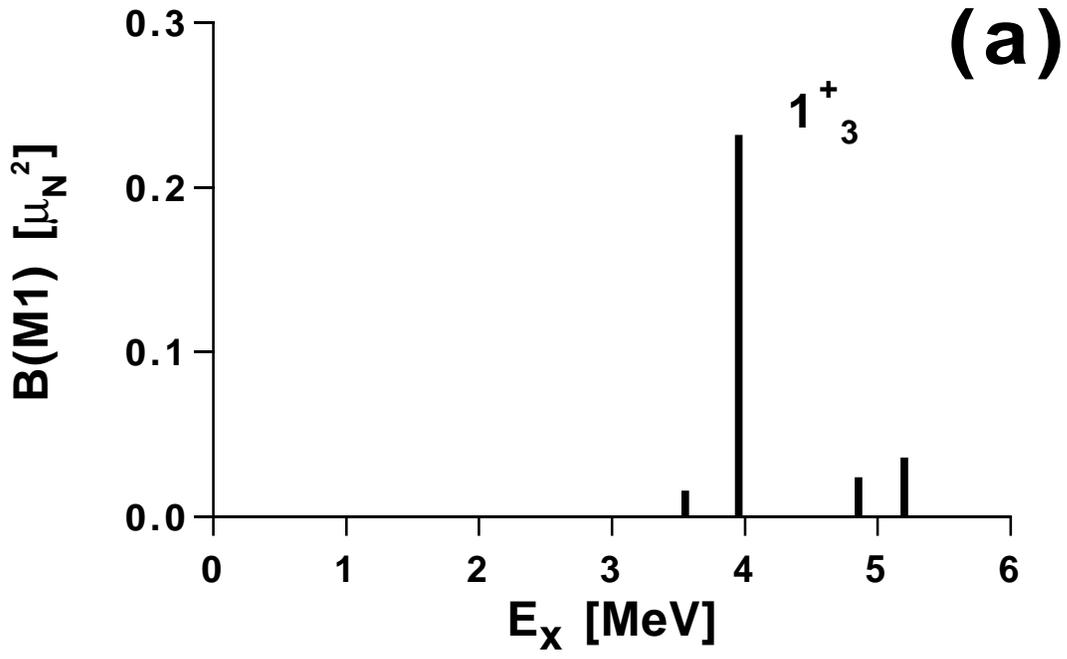
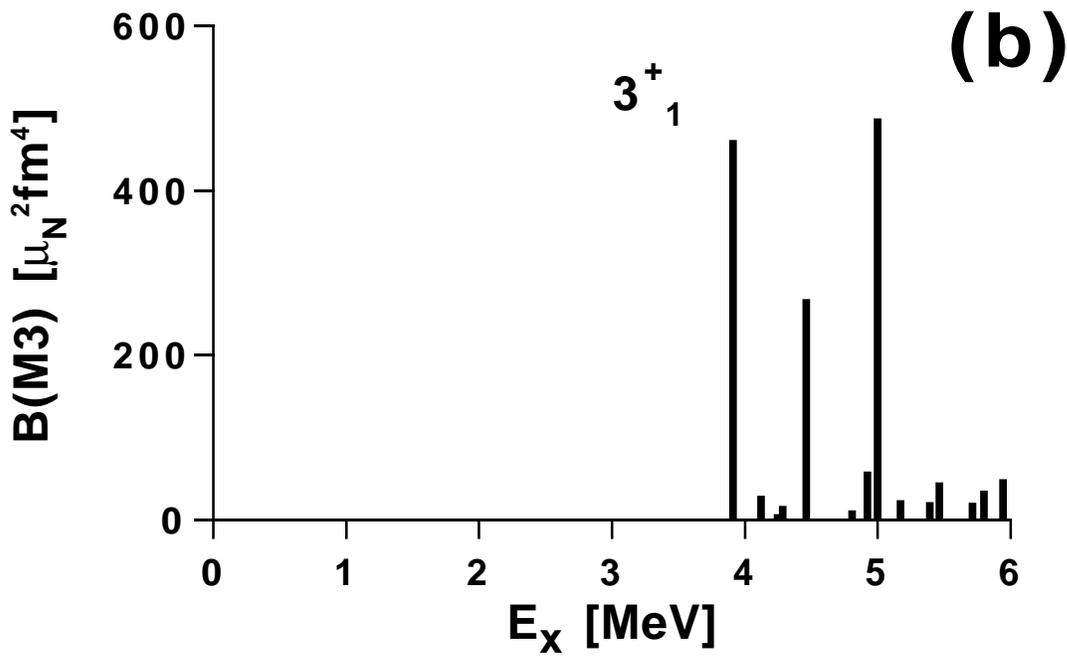
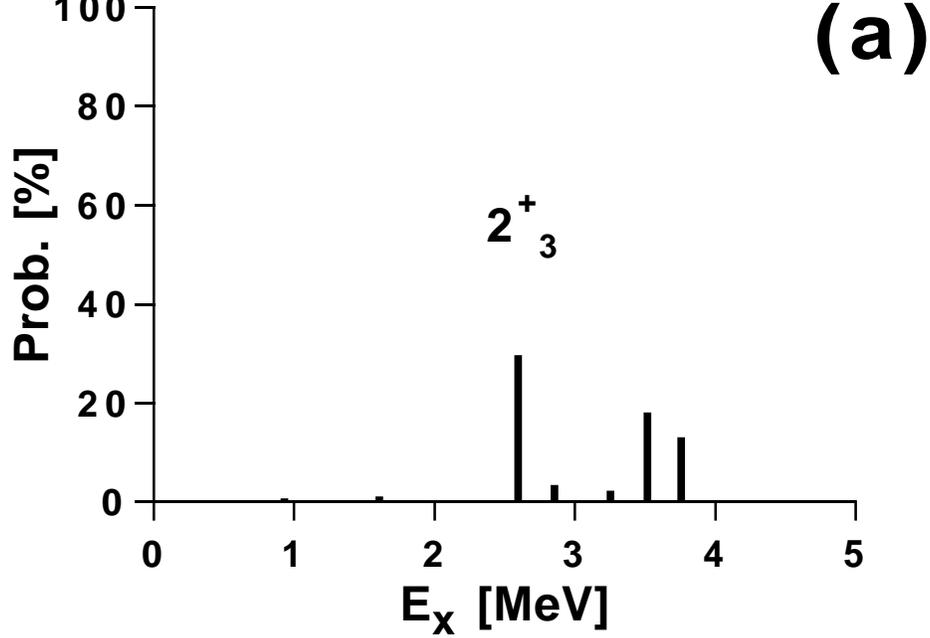
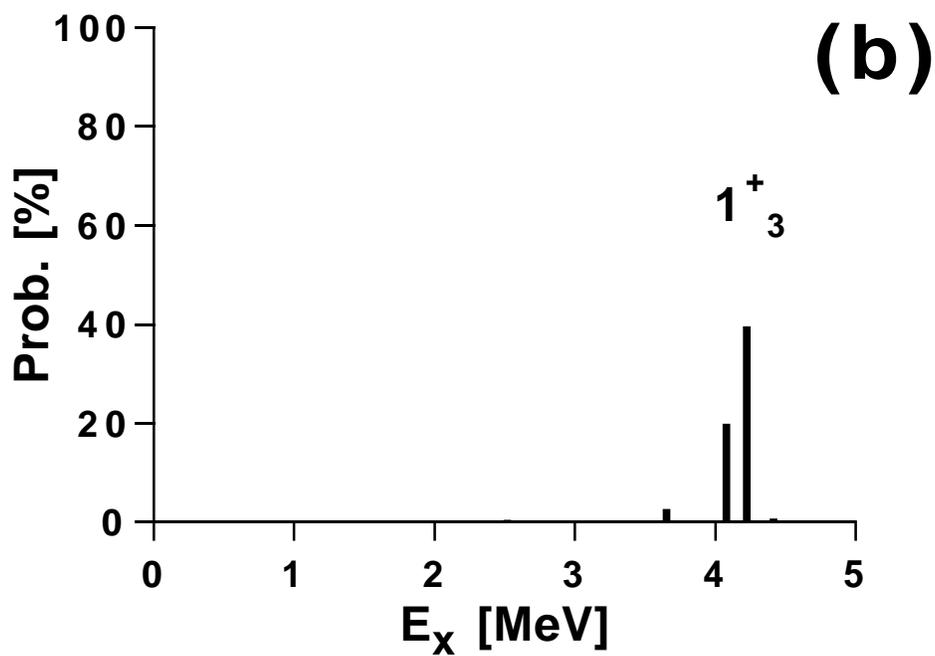
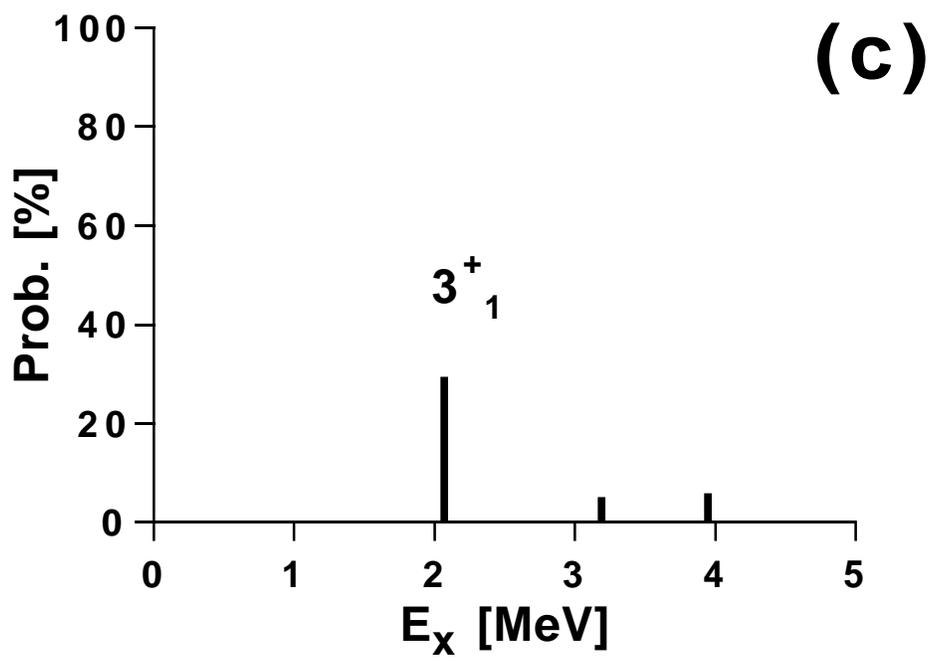
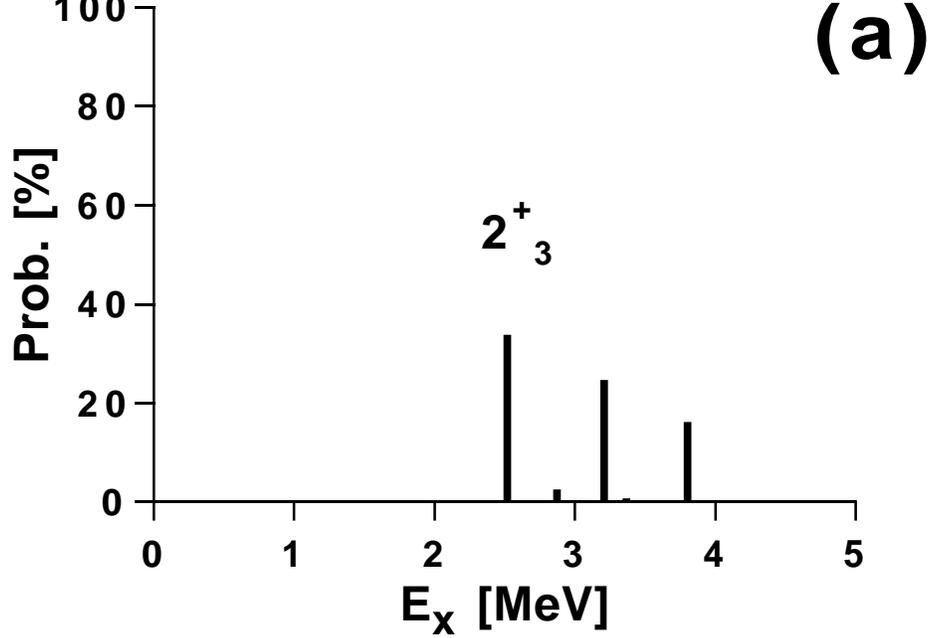
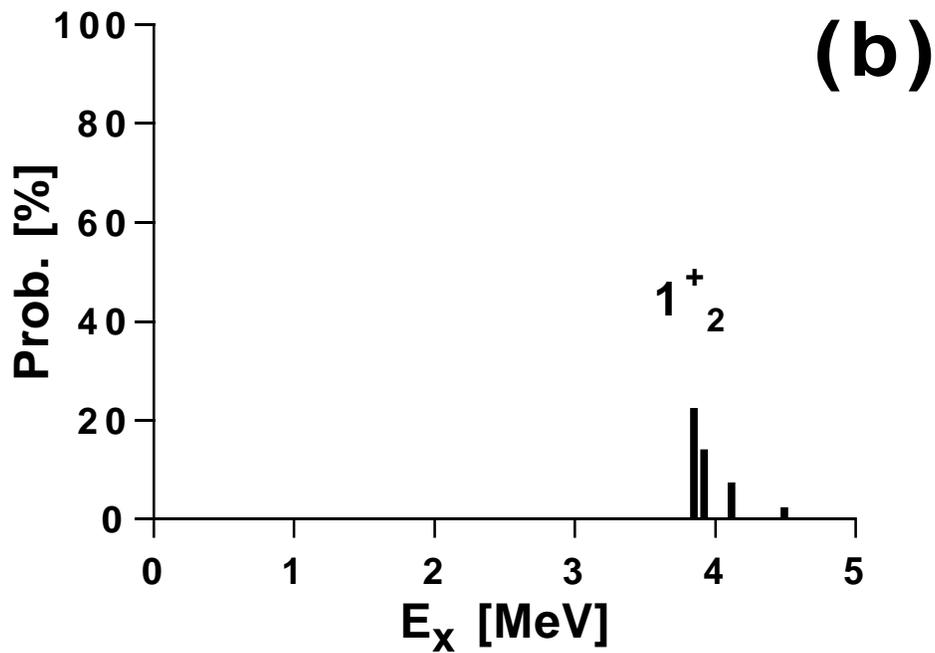
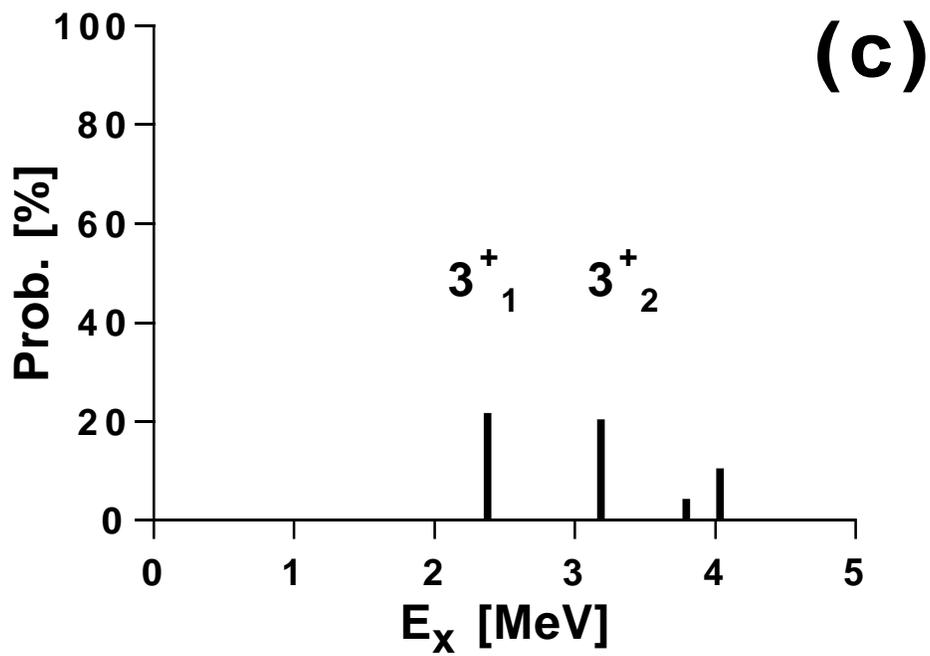
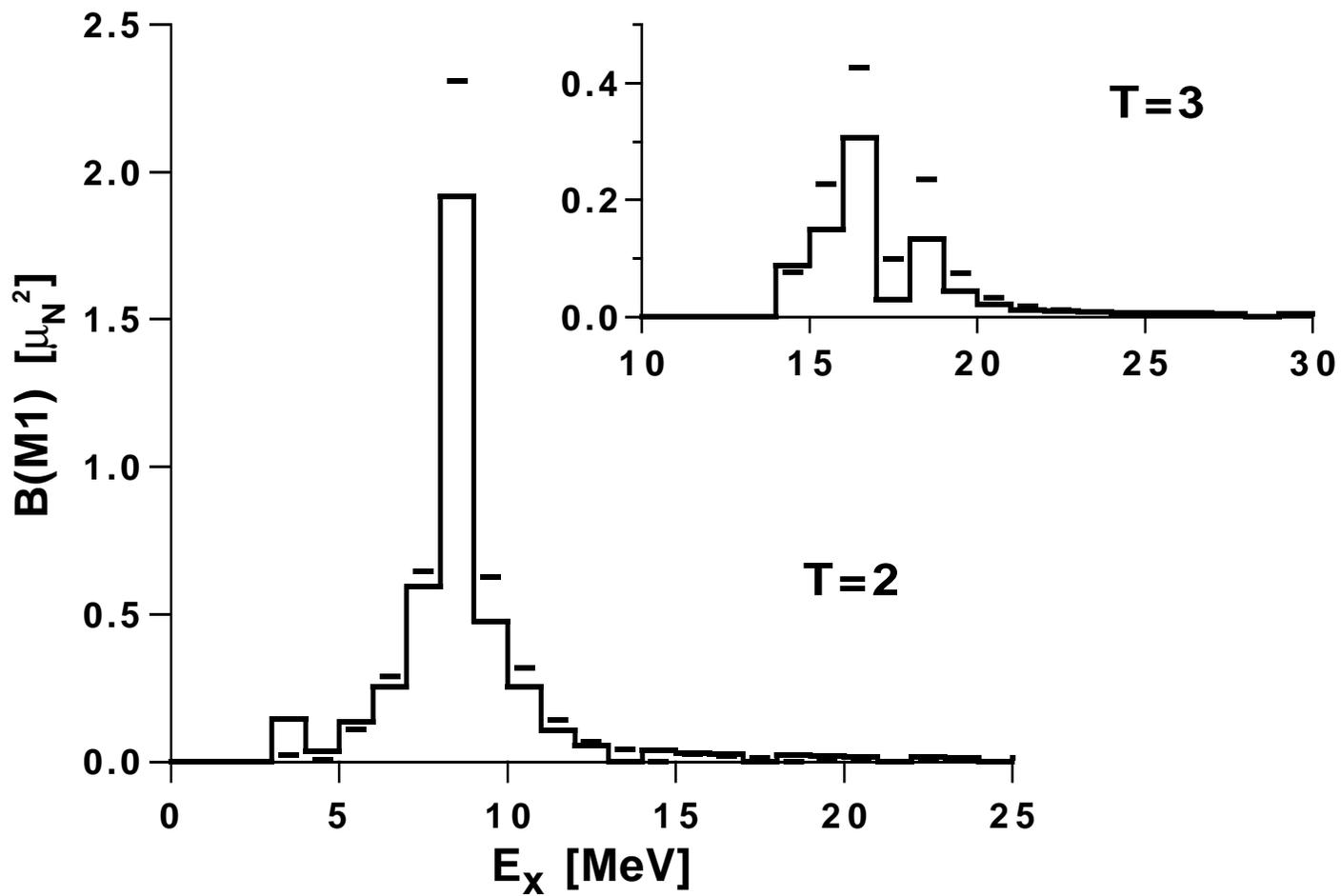
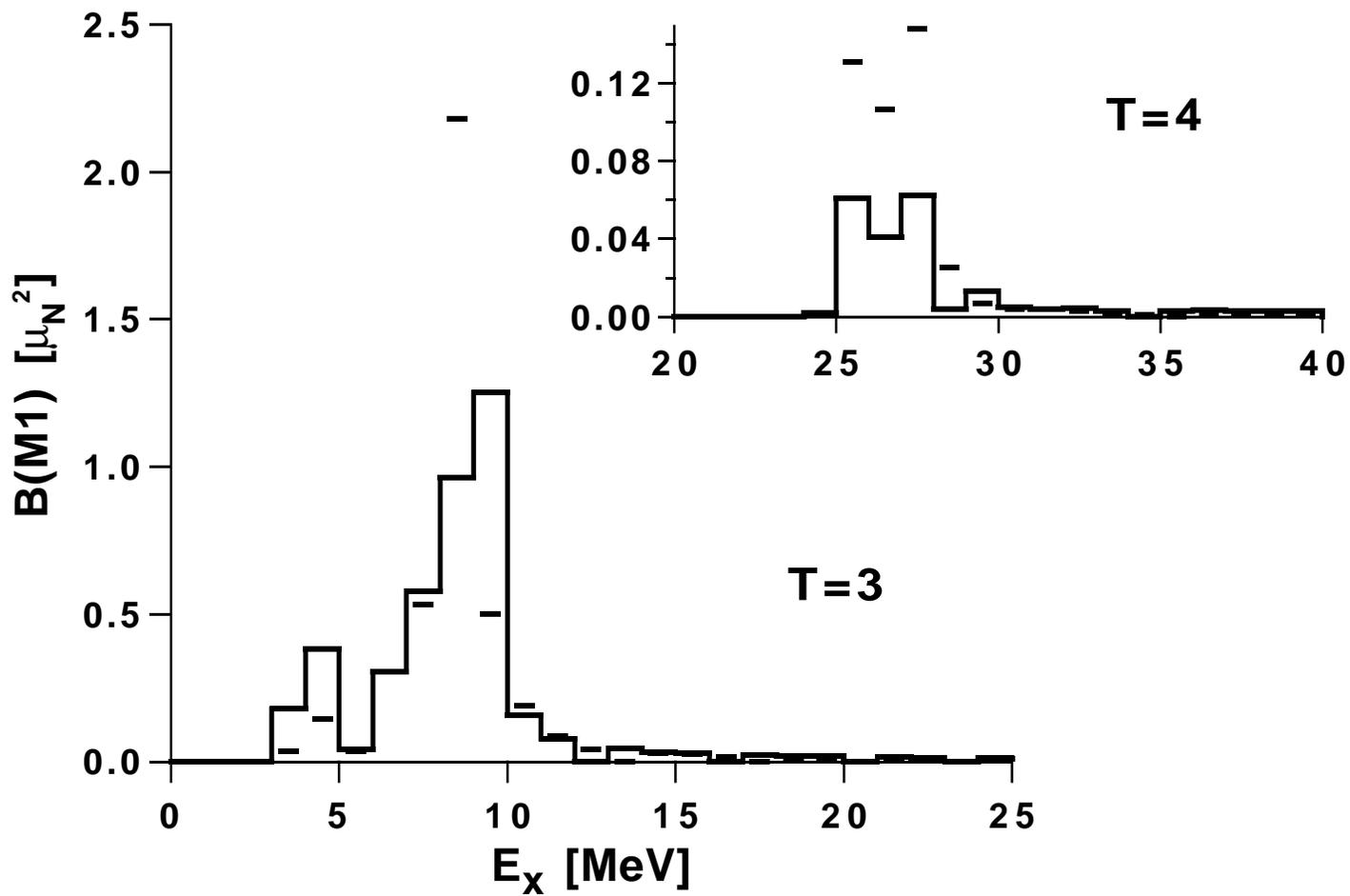
\end{document}